\documentclass[twocolumn,prb,citeautoscript,superscriptaddress,8pt]{revtex4-2}

\usepackage{array}[=2016-10-06]
\usepackage{babel}

\usepackage{graphicx}
\usepackage{gensymb}
\usepackage{color}
\usepackage[dvipsnames]{xcolor}
\usepackage{float}
\usepackage{upgreek}
\usepackage{wrapfig}
\usepackage{bm}
\usepackage{amsmath,amssymb}

\usepackage{multirow}
\usepackage{hyperref}
\usepackage{subcaption}

\newcolumntype{L}{>{\centering\arraybackslash}m{4cm}}
\newcolumntype{J}{>{\centering\arraybackslash}m{6cm}}
\newcolumntype{C}{>{\centering\arraybackslash}m{2cm}}
\newcolumntype{N}{>{\centering\arraybackslash}m{1.7cm}}
\newcolumntype{O}{>{\centering\arraybackslash}m{2.5cm}}
\newcolumntype{M}[1]{>{\centering\arraybackslash}m{#1}}

\newcommand\Tstrut{\rule{0pt}{3ex}}         
\newcommand\Bstrut{\rule[-3ex]{0pt}{0pt}}   
\hypersetup{urlcolor=teal, colorlinks=true, linkcolor=teal, citecolor=teal, linktocpage=true}
\usepackage[toc,page]{appendix}

\begin{document}

\title{Power sensitivity of broadband radiofrequency detectors based on quantum diamond spins}

\author{Nicholas Gillespie}
\affiliation{Department of Physics, School of Science, RMIT University, Melbourne, VIC 3001, Australia}

\author{Christopher T.-K. Lew}
\email[Corresponding author: ]{christopher.tao-kuan.lew@rmit.edu.au}
\affiliation{Department of Physics, School of Science, RMIT University, Melbourne, VIC 3001, Australia}

\author{Ryan Kinsella}
\affiliation{Phasor Quantum Pty Ltd, Level 2/700 Swanston St, Carlton, VIC 3053, Australia} 

\author{Andy Sayers}
\affiliation{Phasor Quantum Pty Ltd, Level 2/700 Swanston St, Carlton, VIC 3053, Australia}
\affiliation{School of Physics, The University of Melbourne, Parkville, VIC 3010, Australia}

\author{Brant Gibson}
\affiliation{Department of Physics, School of Science, RMIT University, Melbourne, VIC 3001, Australia}

\author{David A. Broadway}
\affiliation{Department of Physics, School of Science, RMIT University, Melbourne, VIC 3001, Australia}

\author{Jean-Philippe Tetienne}
\affiliation{Department of Physics, School of Science, RMIT University, Melbourne, VIC 3001, Australia}

\begin{abstract}

Nitrogen-vacancy (NV) centres in diamond can be used to detect radiofrequency (RF) signals through coupling of the RF magnetic field with the NV spins, combined with optical readout of the spin state. The sensitivity of such RF detectors has so far been mainly studied in terms of magnetic field sensitivity, which is relevant when the RF signal is generated by a near-field source. However, for applications where the RF input is delivered externally, a more relevant quantity is the sensitivity in terms of the input RF power. Here we theoretically analyse the power sensitivity of NV-based RF detectors as a function of the RF-spin interface geometry. We derive scaling laws of the power sensitivity for both slope-detection and variance-detection RF sensing protocols, and for various noise regimes. We find that, in most scenarios, the power sensitivity scales inversely with the characteristic physical dimension of the RF-spin interface, for instance the width of a coplanar waveguide or the diameter of a loop antenna. In other words, the smaller the structure and the probed NV volume, the better the power sensitivity, which is contrary to the case of magnetic field sensitivity. Lastly, we numerically estimate that photon shot noise limited sensitivities of $10^{-20}$ W Hz$^{-1}$ (slope) and $10^{-12}$ W Hz$^{-1/2}$ (variance) are achievable. This work lays the groundwork for further optimisation of NV-based RF detectors.

\end{abstract}

\maketitle 

Nitrogen-vacancy (NV) centres in diamond have been researched as radiofrequency (RF) sensors for applications ranging from nanoscale nuclear magnetic resonance  (NMR) spectroscopy to the monitoring of remote RF sources~\cite{rizzato2023review,Du2024review,chen_quantum_2023,shao2016,magaletti_quantum_2022}. Because the NV spins couple to the magnetic field component of the RF field, the sensitivity of NV-based RF detectors is generally quoted in terms of magnetic field, in units of T/$\sqrt{\rm Hz}$. To enable the probing of the broad RF spectrum, various measurement protocols have been developed with the ability to target different frequency sub-bands. For example, RF signals at frequencies close to the NV zero-field splitting of 2.87\,GHz can be detected through direct RF-optical transduction, where the RF field resonant with the NV spin causes a drop in photoluminescence (PL) intensity~\cite{chipaux2015,shao2016}. More advanced protocols applicable for frequencies in the $\sim2$-4\,GHz range include heterodyne methods~\cite{meinel_heterodyne_2021,wang2022}, AC Zeeman measurements~\cite{Ogawa2023}, and differential spin refocusing~\cite{chen2024}. The best sensitivity reported so far in this frequency range is 8.9\,pT/$\sqrt{\rm Hz}$~\cite{wang2022} using the heterodyne protocol. RF signals at lower frequencies from sub-MHz to a few MHz can be detected using Hahn echo ~\cite{wolf2015,barry2024} and higher-order pulsed dynamical decoupling protocols~\cite{delange2011,Glenn2018}. The range of these protocols can be extended to higher frequencies up to $\sim100$\,MHz using continuous dynamical decoupling schemes~\cite{loretz_radio-frequency_2013,Stark2017,hermann_extending_2024,louzon_robust_2025}, and even to any arbitrary frequency using quantum frequency mixing~\cite{Wang2022_mixer,yin_high-resolution_2025}. Example reported magnetic sensitivities are 210\,fT/$\sqrt{\rm Hz}$ at 6.4\,kHz with Hahn echo~\cite{barry2024}, 32\,pT/$\sqrt{\rm Hz}$ at 3.7\,MHz with pulsed dynamical decoupling~\cite{Glenn2018}, 70\,fT/$\sqrt{\rm Hz}$ at 0.35\,MHz, enhanced through magnetic flux concentration \cite{silani_nuclear_2023}, and 1-100\,nT/$\sqrt{\rm Hz}$ over the 10 MHz to 4 GHz range with quantum frequency mixing~\cite{yin_high-resolution_2025}.  

\begin{figure*}[]
    \centering
    \includegraphics[scale =1.7]{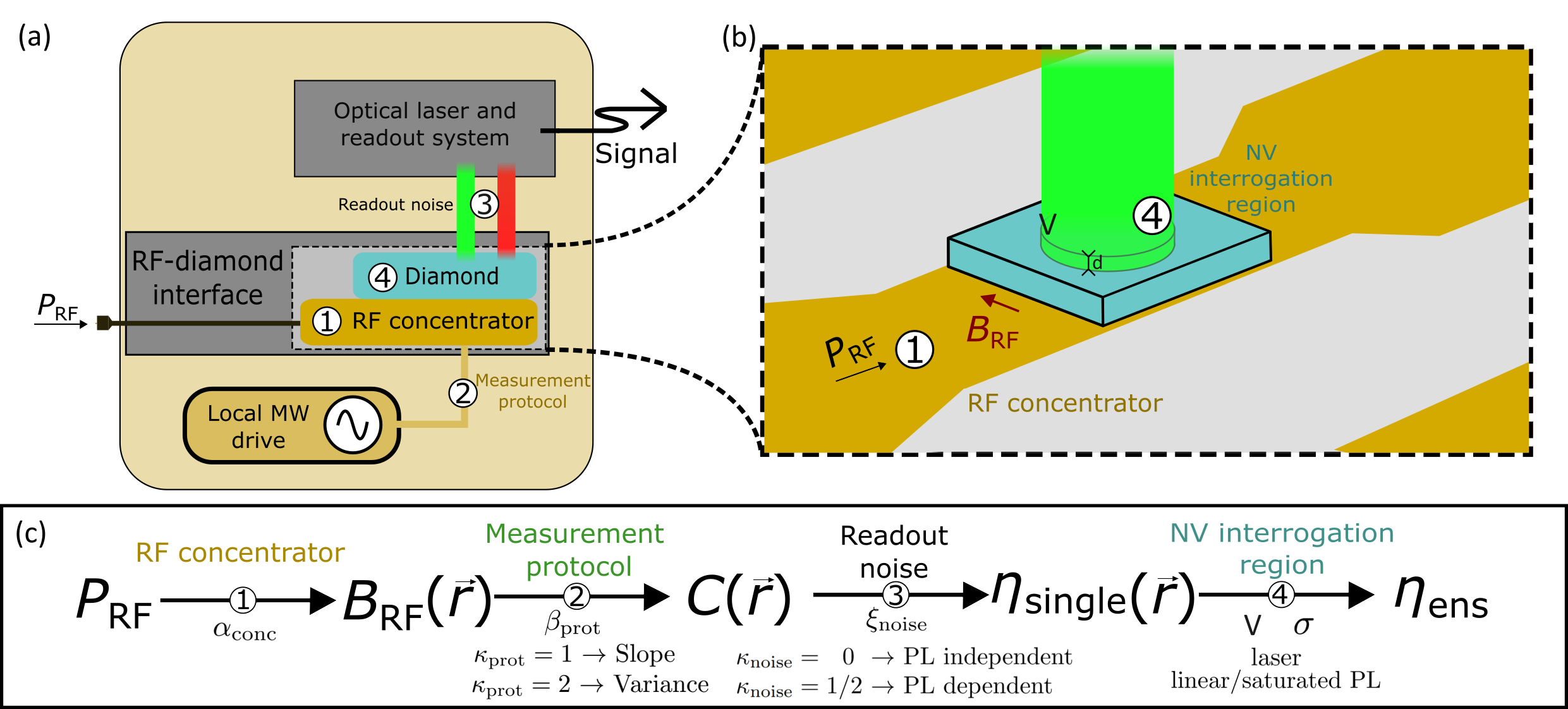}
    \caption{(a) Generalised structure of an NV-based RF detector, where the RF-diamond interface is specific to the RF concentrator geometry. (b) Example of a common RF concentrator, the coplanar waveguide. (c) Workflow of logic for defining the power sensitivity relationships. Each step details constants (defined in the main text) coupling the two stages. Each transition, referred to by number, is visually allocated to system diagrams in (a,b).}
    \label{BigPicture}
\end{figure*}

When the RF signal to be detected originates from a remote source, the RF magnetic field experienced by the NV spins depends on the method used to deliver the input RF signal. In this work, we consider the generic scenario where the input RF signal enters the detector through a coaxial waveguide which feeds a broadband RF concentrator structure (e.g. a waveguide, or antenna) interfacing with the NV spins (Fig. \ref{BigPicture}(a)). The role of the RF concentrator is to concentrate the input RF power ($P_{\rm RF}$) into a small volume ($V$) occupied by NV centres, which then experience an RF magnetic field, $B_{\rm RF}$. In this scenario, the relevant sensitivity of the device is in terms of the input power $P_{\rm RF}$ rather than the magnetic field. That is, we are interested in the minimum input power $\delta P_{\rm RF}$ that can be detected in a given amount of time. Little attention has so far been paid to optimising this power sensitivity. For example, some of the best magnetic sensitivities reported to date were obtained by maximising the volume of NVs being probed ($V$), but this necessitated a large RF concentrator (typically a simple loop antenna was employed) with a low conversion ratio from $P_{\rm RF}$ to $B_{\rm RF}$~\cite{wang2022,barry2024,Glenn2018} which will negatively impact the power sensitivity. In this context, it is not a priori obvious whether it is preferable to maximise $B_{\rm RF}$ or $V$, which impose opposite requirements on the size of the RF concentrator (which should be minimised or maximised, respectively).

In this paper, we study the power sensitivity of NV-based weak-field RF detectors through analytical derivations complemented by numerical validation. We analyse in particular the sensitivity scaling with the system's physical dimensions for two common types of broadband RF concentrators: coplanar waveguide and loop antenna. We consider two classes of RF sensing protocols representative of most possible sensing regimes: variance-detection and slope-detection protocols. We also consider different noise and laser saturation regimes. Our analysis shows that, in the case of field-variance RF detection in the photon shot noise limit, the power sensitivity improves as the physical dimension of the RF concentrator (and hence the volume of probed NVs) is reduced, which is contrary to the case of magnetic field sensitivity. We also provide numerical estimates of the power sensitivity that may be realistically achieved. 
This work will guide future work on optimising and implementing NV-based RF detectors.  

\section{General framework}

The generalised RF detection system can be seen in Fig. \ref{BigPicture}(a)-(b), which involves a coaxial cable (connected upstream to a source e.g. a receiver antenna, transmitter or device under test) carrying the RF signal of interest and delivering it to the diamond through an RF concentrator. 
In this section, we derive general equations for the power sensitivity of the device as a function of factors characterising the RF concentrator (1), the RF sensing protocol (2),  the measurement noise (3), and the NV interrogation volume (4).
We will later specify some of these factors for different types of RF concentrators. A summary of the symbols used in this paper can be found in Appendix \ref{sec:SumTable}, Table \ref{Appx:SumTable}. 

The workflow for deriving the power sensitivity is shown in Fig. \ref{BigPicture}(c). In step 1, the input RF power $P_{\rm RF}$ is converted to an RF magnetic field distribution $B_{\rm RF}(\vec{r})$. Here, $B_{\rm RF}$ is the RMS amplitude of the field component the NV spins are sensitive to for the sensing protocol considered, which is either parallel or transverse to the NV axis~\cite{Degen2017}. This distribution depends on the details of the RF concentrator but can generally be parametrised as $B_{\rm RF}(\vec{r})=\alpha_{\rm conc}(\vec{r})\sqrt{P_{\rm RF}}$ where $\alpha_{\rm conc}(\vec{r})$ characterises the RF concentrator. Typically, $\alpha_{\rm conc}(\vec{r})$ will be maximum in some region near the RF concentrator (where the diamond will be positioned), and decay outside this region.

In step 2, the RF magnetic field $B_{\rm RF}(\vec{r})$ at a given position $\vec{r}$ induces a relative PL change (or contrast) in the NV at this location, $C(\vec{r})$. This contrast depends on the RF sensing protocol employed but can be generally parametrised as $C(\vec{r})=[\beta_{\rm prot}B_{\rm RF}(\vec{r})]^{\kappa_{\rm prot}}$ where $\beta_{\rm prot}$ characterises the protocol. Here we assume the weak-field regime where the RF field induces only small rotations around the Bloch sphere~\cite{Degen2017}. The exponent $\kappa_{\rm prot}$ can be either 1 or 2 depending on the protocol due to the small-angle approximations. The case $\kappa_{\rm prot}=2$ corresponds to the class of protocols known as variance-detection protocols, $C\propto B_{\rm RF}^2$, whereas $\kappa_{\rm prot}=1$ corresponds to slope-detection protocols, $C\propto B_{\rm RF}$~\cite{Degen2017}. For example, dynamical decoupling methods can be switched from variance- to slope-detection by simply changing the phase of the final $\pi/2$ projection pulse from 0 to $\pi$/2 in the NV driving sequence~\cite{Laraoui2011,rizzato_extending_2023}. Other protocols such as direct RF-optical transduction are only capable of variance detection~\cite{shao2016}. We note that most RF sensing protocols require a microwave driving field at the NV Larmor frequency (see Fig. \ref{BigPicture}(a)), which may be delivered using the same RF concentrator or using a separate structure such as a loop antenna. We ignore this requirement in our analysis and only consider the unknown RF signal to be measured.   

In step 3 of the workflow, the contrast $C(\vec{r})$ is converted to a power sensitivity for a single NV at this location, $\eta_{\rm single}(\vec{r})$, which depends on the measured photon rate from the NV and the measurement noise. To derive $\eta_{\rm single}(\vec{r})$, we first express the signal-to-noise ratio ${\rm SNR}=S/N$. The signal is given by $S=CI_{\rm PL}t$, where $I_{\rm PL}$ is the collected photon rate (taking into account the duty cycle of the protocol considered) and $t$ the total measurement time. Meanwhile, the noise can generally be modelled as $N=\xi_{\rm noise}I_{\rm PL}^{\kappa_{\rm noise}}\sqrt{t}$. An exponent $\kappa_{\rm noise}=0.5$ is the most common case where the noise scales with the PL intensity. The pre-factor is $\xi_{\rm noise}=1$ in the photon shot noise limit, but may be larger in practice. An exponent $\kappa_{\rm noise}=0$ corresponds to the PL-independent limit which may apply when the instrumentation reaches its noise floor. Here we consider these two cases independently with $\kappa_{\rm noise}=0$ or $0.5$.

Combining all the above formulas, we obtain a general expression for the SNR of a single NV:
\begin{equation}
    \label{SNR}
    {\rm SNR}_{\rm single} = \frac{(\alpha_{\rm conc}\beta_{\rm prot})^{\kappa_{\rm prot}}}{\xi_{\rm noise}}I_{\rm PL}^{1-\kappa_{\rm noise}}\sqrt{t}P_{\rm RF}^{\kappa_{\rm prot}/2}~.
\end{equation}
The power sensitivity, $\eta_{\rm single}$, is defined as the minimum RF power $\delta P_{\rm RF}$ leading to an SNR of unity, with the appropriate time normalisation. For variance-detection protocols ($\kappa_{\rm prot}=2$), we obtain
\begin{equation}
    \label{single_var}
    \eta_{\rm single,var} = \delta P_{\rm RF}\sqrt{t} = \frac{\xi_{\rm noise}}{\alpha_{\rm conc}^2\beta_{\rm prot}^2I_{\rm PL}^{1-\kappa_{\rm noise}}}~,
\end{equation}
in units of W/$\sqrt{\rm Hz}$. For slope-detection protocols ($\kappa_{\rm prot}=1$), we obtain
\begin{equation}
    \label{single_slope}
    \eta_{\rm single,slope} = \delta P_{\rm RF}t = \left[\frac{\xi_{\rm noise}}{\alpha_{\rm conc}\beta_{\rm prot}I_{\rm PL}^{1-\kappa_{\rm noise}}}\right]^2~,
\end{equation}
in units of W/Hz. For convenience these equations can be combined into a single expression,
\begin{equation}
    \label{single}
    \eta_{\rm single} = \delta P_{\rm RF}t^{1/\kappa_{\rm prot}} = \left[\frac{\xi_{\rm noise}}{(\alpha_{\rm conc}\beta_{\rm prot})^{\kappa_{\rm prot}}I_{\rm PL}^{1-\kappa_{\rm noise}}}\right]^{2/{\kappa_{\rm prot}}}~.
\end{equation}

Finally, in step 4 of the workflow, we extend the power sensitivity of a single NV at location $\vec{r}$ to an ensemble-averaged sensitivity $\eta_{\rm ens}$, where the PL is collected from a given volume of NVs. For simplicity, we assume that the NV density and laser intensity are uniform over the collected volume $V$, and that each NV in the ensemble contributes equally to the total measured photon rate. As such, we can characterise the NV ensemble by a quantity $\rho_{\rm PL}$ (in units of Hz/m$^3$) which is the PL rate measured per unit volume (within the probed volume) such that the total measured PL rate is $I_{\rm PL} = \int_V \rho_{\rm PL}{\rm d}^3\vec{r}=\rho_{\rm PL}V$. The SNR for the ensemble measurement becomes:
\begin{align}
    \label{SNR_ens}
    {\rm SNR}_{\rm ens} & =  \int_V\alpha_{\rm conc}^{\kappa_{\rm prot}}(\vec{r}){\rm d}^3\vec{r}\frac{\beta_{\rm prot}^{\kappa_{\rm prot}}\rho_{\rm PL}^{1-\kappa_{\rm noise}}}{\xi_{\rm noise}V^{\kappa_{\rm noise}}}\sqrt{t}P_{\rm RF}^{\kappa_{\rm prot}/2} \nonumber \\
        & = \langle\alpha_{\rm conc}^{\kappa_{\rm prot}}\rangle_V\frac{\beta_{\rm prot}^{\kappa_{\rm prot}}(\rho_{\rm PL}V)^{1-\kappa_{\rm noise}}}{\xi_{\rm noise}}\sqrt{t}P_{\rm RF}^{\kappa_{\rm prot}/2}~,
\end{align}
where $\langle x\rangle_V$ denotes the spatial average over the probed volume.    
The ensemble sensitivity is therefore
\begin{equation}
    \label{ens}
    \eta_{\rm ens} = \left[\frac{\xi_{\rm noise}}{\langle\alpha_{\rm conc}^{\kappa_{\rm prot}}\rangle_V\beta_{\rm prot}^{\kappa_{\rm prot}}(\rho_{\rm PL}V)^{1-\kappa_{\rm noise}}}\right]^{2/{\kappa_{\rm prot}}}~.
\end{equation}

We now seek to isolate the geometry-dependent terms in Eq. \ref{ens}. Apart from $\langle\alpha_{\rm conc}^{\kappa_{\rm prot}}\rangle_V$ and $V$, the other factor that may depend on geometry is $\rho_{\rm PL}$, the PL rate per unit volume. We can express it in the form of a saturation law 
\begin{equation}
    \label{sat}
    \rho_{\rm PL}=\sigma_{\rm NV}\rho_{\rm NV}\frac{1}{1+\frac{AI_{\rm sat}}{P_{\rm laser}}}~,
\end{equation}
where $\rho_{\rm NV}$ is the NV density, $P_{\rm laser}$ is the laser power, $A$ the cross-sectional area of the laser beam, and $I_{\rm sat}\sim1$\,mW$/\mu$m$^2$ is the NV saturation intensity~\cite{Manson2006}. The factor $\sigma_{\rm NV}$ is the PL rate measured for a single NV at saturation, accounting for the PL collection efficiency and the duty cycle of the sensing protocol. 

Here we can distinguish two regimes. For sufficiently small probe volumes $V$, we can assume that there is sufficient laser power available to reach optical saturation of the NVs. In this case, $\rho_{\rm PL}=\sigma_{\rm NV}\rho_{\rm NV}$ is a constant independent of geometry. We can define a figure of merit that captures the geometry-dependent factors in the sensitivity equation,
\begin{equation}
    \label{FOM1}
    {\rm FoM}_{\rm sat} = \langle\alpha_{\rm conc}^{\kappa_{\rm prot}}\rangle_V V^{1-\kappa_{\rm noise}}~.
\end{equation}
If the laser power is fixed and always insufficient to reach saturation, then we have $\rho_{\rm PL}\propto1/A$ and the relevant figure of merit for this regime (thereafter referred to as the linear regime) is
\begin{equation}
    \label{FOM2}
    {\rm FoM}_{\rm lin} = \langle\alpha_{\rm conc}^{\kappa_{\rm prot}}\rangle_V \left(\frac{V}{A}\right)^{1-\kappa_{\rm noise}} = \langle\alpha_{\rm conc}^{\kappa_{\rm prot}}\rangle_V d^{1-\kappa_{\rm noise}}
\end{equation}
where $d$ is the distance traversed by the laser beam across the probe volume $V$ (see Fig. \ref{BigPicture}(b)).
In both regimes, the sensitivity scales as 
\begin{equation}
    \label{sensFOM}
    \eta_{\rm ens}\propto{\rm FoM}^{-2/{\kappa_{\rm prot}}}.
\end{equation}
Thus, the geometry of the problem, namely the design and dimensions of the RF concentrator, the choice of the probe volume, and the direction of the laser beam, should be optimised to maximise the relevant figure of merit ${\rm FoM}_{\rm sat}$ or ${\rm FoM}_{\rm lin}$.

\section{Application to specific RF concentrator geometries}

We now apply this general framework to specific RF concentrator designs and derive scaling laws for the power sensitivity as a function of key geometrical parameters. Different regimes of protocols (variance or slope detection), noise (PL-dependent or independent), laser intensity (saturating the NV PL or power limited), and laser beam direction, are analysed. 

\subsection{Coplanar waveguide}

\subsubsection{Model}

\begin{figure*}
    \centering
    \includegraphics[width=1\linewidth]{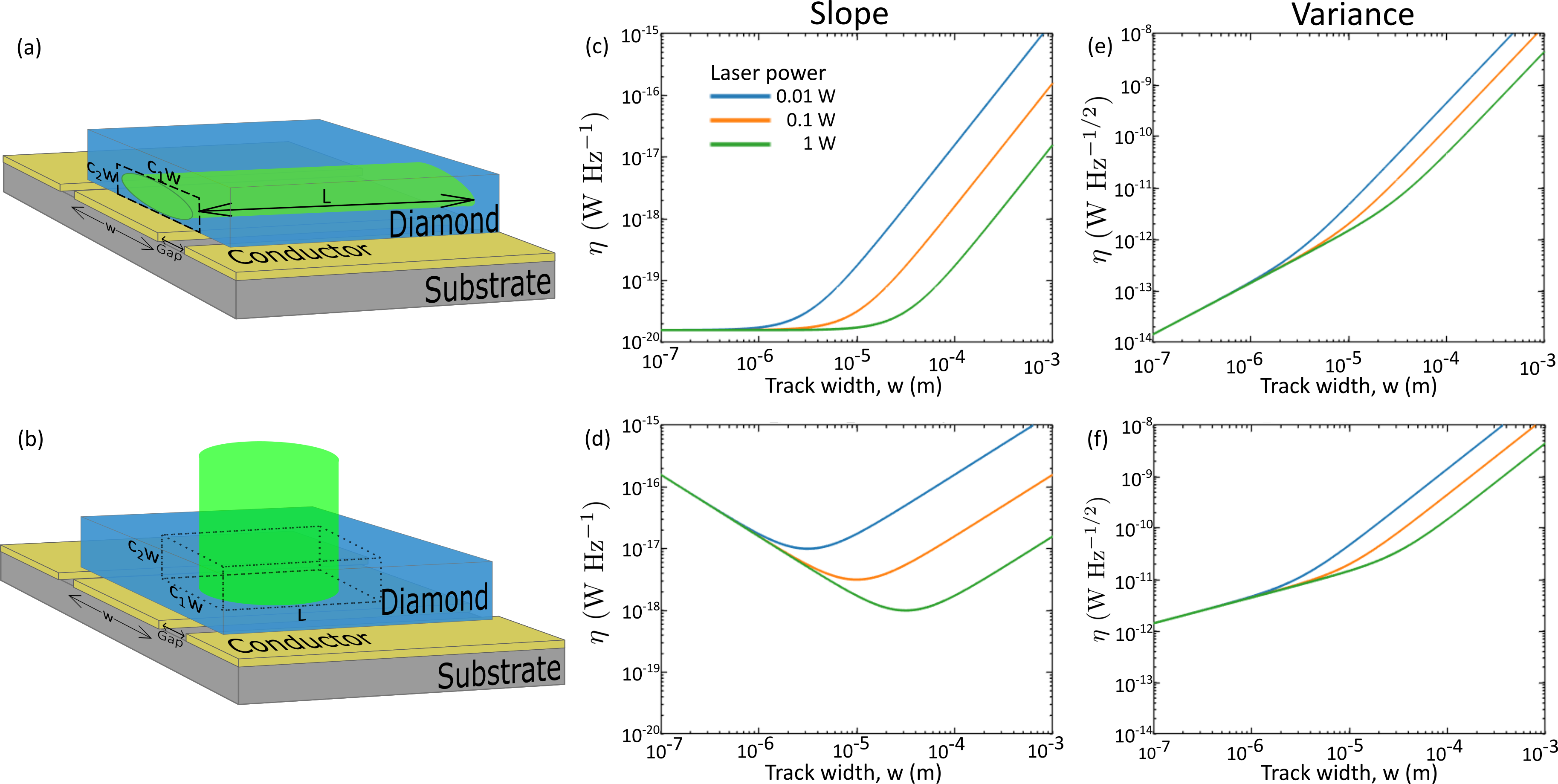}
    \caption{(a)-(b) Two possible interrogation geometries for the coplanar waveguide, with a laser beam parallel (a) or perpendicular (b) to the waveguide. (c)-(f) Power sensitivity of the slope (c)-(d) and variance (e)-(f) protocols as a function of the width of the CPW, with a laser beam parallel (c),(e) or perpendicular (d),(f) to the waveguide. The different curves correspond to different laser powers. The sensitivity is calculated by substituting Eq. \ref{sat} into Eq. \ref{ens_CPW}. Only the PL-dependent noise regime is considered ($\kappa_{\rm noise}=1/2$). The various constants are taken to be (see justification in Sec. \ref{sec:num}) $\beta_{\rm{prot}}=10^{4}~{\rm T}^{-1}$, $\rho_{\rm{NV}} = 8\times10^{23}~{\rm m}^{-3}$, $I_{\rm sat}=1$\,mW$/\mu$m$^2$, $\sigma_{\rm NV}=10^{5}~{\rm s}^{-1}$, $Z = 50~\Omega$, $L = 1$~mm (or $L=w$ in the case of perpendicular excitation), and $\zeta=\xi_{\rm{noise}}=c_1=c_2=1$.
    }
    \label{fig:CPW}
\end{figure*}

The coplanar waveguide (CPW) provides an easy to fabricate, broadband solution for the local delivery of RF signals to the NV spins. Moreover, it allows the field-to-power ratio (i.e. the factor $\alpha_{\rm conc}$) to be increased by simply reducing the transverse size of the CPW by tapering from a wider source \cite{jia_ultra-broadband_2018}. The CPW can be approximately characterised by two independent parameters, e.g. the track width $w$ and the impedance $Z$ (which together dictate the gap between the central track and the return conductors). 
The diamond is placed on top of the CPW and we assume that only the NVs within an area of interest above the central track are measured (see Fig. \ref{fig:CPW}(a)-(b)). At the surface of the track, the RF magnetic field has a root-mean-square (RMS) amplitude of~\cite{Tetienne2019} 
\begin{equation}
    \label{B_CPW}
    B_{\rm RF} \approx \frac{\mu_0I_{\rm RF}}{2w} = \frac{\mu_0}{2w\sqrt{Z}}\sqrt{P_{RF}}
\end{equation}
where we used the lossless transmission line relation $I_{\rm RF}=\sqrt{P_{\rm RF}/Z}$. 
The magnetic field decays as one moves away from the surface and centre of the track, reducing the local value of $\alpha_{\rm conc}$. As a result, there will be an optimum choice for the probe volume $V$ that maximises the figure of merit in Eq. \ref{FOM1} or \ref{FOM2}. In practice, this volume should be numerically optimised given experimental constraints, for instance the shape of $V$ is constrained by the diamond and optical design. 

To progress our analysis, we consider the volume $V$ to be a parallelepiped with rectangular cross section of width $c_1w$, height $c_2w$, and length $L$ along the CPW. We also make the assumption (validated numerically in Appendix \ref{app:CPW num}) that the optimum ratios $c_{1,2}$ are independent of $w$. Moreover, the average magnetic field within this optimum volume will be below but of the order of the maximum value given by Eq.~\ref{B_CPW}, such that we can express the average field-to-power ratio as 
\begin{equation}
    \label{alpha_opt}
    \langle\alpha_{\rm conc}^{\kappa_{\rm prot}}\rangle_V\approx \zeta\left[ \frac{\mu_0}{2w\sqrt{Z}}\right]^{\kappa_{\rm prot}}~,
\end{equation}
where $\zeta$ is the ratio between the approximated field and the average field over the probed region (we omit the $\kappa_{\rm prot}$ dependence for conciseness). Note that we have assumed here that the NV spins have the correct orientation to maximise coupling to the in-plane magnetic field for the sensing protocol considered.

Substituting Eq. \ref{alpha_opt} and the volume $V=c_1c_2w^2L$ into Eq. \ref{ens}, we obtain the power sensitivity equation for the CPW,
\begin{equation}
    \label{ens_CPW}
    \eta_{\rm CPW} = \left[\frac{\xi_{\rm noise}}{\zeta\left[ \frac{\mu_0}{2w\sqrt{Z}}\right]^{\kappa_{\rm prot}}\beta_{\rm prot}^{\kappa_{\rm prot}}(\rho_{\rm PL}c_1c_2w^2L)^{1-\kappa_{\rm noise}}}\right]^{2/{\kappa_{\rm prot}}}~.
\end{equation}

\subsubsection{Scaling laws}

From Eq. \ref{ens_CPW}, we can derive scaling laws for the sensitivity as a function of the geometrical parameters $w$ and $L$, in the saturated PL regime and in the linear regime. Two different laser beam geometries are considered, either parallel to the CPW (Fig.~\ref{fig:CPW}(a)) or perpendicular to it (Fig.~\ref{fig:CPW}(b)).

In the saturated regime, we assume there is sufficient laser power available to reach NV optical saturation regardless of the volume $V=c_1c_2w^2L$. In this case, $\rho_{\rm PL}=\sigma_{\rm NV}\rho_{\rm NV}$ is independent of geometrical parameters and we deduce the following scaling law,
\begin{equation}
    \label{scaling_CPW1}
    \eta_{\rm CPW} \propto w^{2-4\frac{1-\kappa_{\rm noise}}{\kappa_{\rm prot}}} L^{-2\frac{1-\kappa_{\rm noise}}{\kappa_{\rm prot}}} ~.
\end{equation}
For the parallel beam geometry (Fig.~\ref{fig:CPW}(a)), $L$ and $w$ can be varied independently. In practice, this may be achieved by patterning an optical waveguide into the diamond~\cite{Lenzini2018}, where $L$ would be the length of the diamond of the order of 1\,mm. For the perpendicular beam geometry (Fig.~\ref{fig:CPW}(b)), setting $L$ independently of $w$ would be impractical in practice. Instead we assume $L=w$ to maintain a beam aspect ratio of order unity. The resulting explicit scaling laws for the different protocols and noise regimes and the two laser beam geometries are given in Table \ref{table:CPW}. 

In the linear regime, $\rho_{\rm PL}=\sigma_{\rm NV}\rho_{\rm NV}\frac{P_{\rm laser}}{AI_{\rm sat}}$ remains far from saturation given a fixed laser power. For the parallel beam geometry, we have $A=c_1c_2w^2$ which leads to the sensitivity scaling as
\begin{equation}
    \label{scaling_CPW2}
    \eta_{\rm CPW} \propto w^{2} L^{-2\frac{1-\kappa_{\rm noise}}{\kappa_{\rm prot}}} ~.
\end{equation}
If instead the laser beam comes from the top with a beam size $A=c_1wL$, we obtain
\begin{equation}
    \label{scaling_CPW3}
    \eta_{\rm CPW} \propto w^{2-2\frac{1-\kappa_{\rm noise}}{\kappa_{\rm prot}}} L^0 ~.
\end{equation}
We note that the sensitivity is now independent of $L$ and so we do not have to assume a beam aspect ratio of unity. This invariance with beam ratio in the linear regime reflects the constancy of collected PL due to the inverse relationship between collection area and PL density.
The explicit scaling laws for the different scenarios considered are given in Table \ref{table:CPW}.

\subsubsection{PL-dependent noise regime}

We now discuss these scaling laws and their implications for optimising the power sensitivity. We focus on the PL-dependent noise regime ($\kappa_{\rm noise}=1/2$) as this is the most interesting scenario encompassing the photon shot noise limit.

In Fig.~\ref{fig:CPW}(c-f), we plot $\eta_{\rm CPW}$ from Eq. \ref{ens_CPW} as a function of the track width $w$ when substituting the general saturation law for $\rho_{\rm PL}$. Fig.~\ref{fig:CPW}(c)-(d) show the slope and variance detection cases, respectively, for the parallel beam geometry, while Fig.~\ref{fig:CPW}(e)-(f) show the perpendicular beam case. The different curves correspond to different values of the laser power $P_{\rm laser}$. In each graph in Fig.~\ref{fig:CPW}(c)-(f), we can distinguish two regimes: the linear PL regime for large $w$, and the saturated PL regime for small $w$. This allows us to associate the trends in each regime with the scaling laws of Table \ref{table:CPW}. The crossover between the linear and saturated regimes depends on the laser power but typically occurs at a width of the order of $w\sim10\,\mu$m.

In all cases plotted in Fig.~\ref{fig:CPW}(c)-(f), we see that starting from a large $w$, the sensitivity improves as $w$ is reduced, with a dependence ranging from linear ($w^1$) to quadratic ($w^2$) as it approaches saturation. Once the probe volume becomes small enough to saturate the NVs for a given laser power, we observe a further improvement in sensitivity for variance detection as $w$ is further reduced, albeit with a reduction in slope compared to the linear regime (e.g. going from a $w^2$ to $w^1$ scaling for the parallel beam). For slope detection, however, we observe either a plateau in sensitivity for a parallel beam, or an inversion for a perpendicular beam indicating there is an optimum width that is laser power dependent. These trends contrast with magnetic field sensitivities, which are generally improved or maintained when the probe volume is increased~\cite{barry2020}. The difference is of course due to the fact that when one considers power sensitivity, the magnetic field becomes a variable that increases with decreasing system's size instead of being constant.  

\begin{table*}
    \begin{tabular}{ |c|c|c|c|c|c| } 
         \hline
        \multirow{4}{8em}{\centering Noise regime} & \multirow{4}{8em}{\centering Laser \\ geometry} &   \multicolumn{4}{c|}{$\eta_{CPW} \propto$}   \\ \cline{3-6}  &  & \multicolumn{2}{c|}{\centering Slope,  $\kappa_{\rm prot} = 1$}  & \multicolumn{2}{c|}{\centering Variance, $\kappa_{\rm prot} = 2$ } \\ \cline{3-6} &  &   \multirow{2}{8em}{\centering Saturated PL} &  \multirow{2}{8em}{\centering Linear PL} &  \multirow{2}{8em}{\centering Saturated PL}  & \multirow{2}{8em}{\centering Linear PL}   \\ &  &  &  & &  \\
        \hline
    
        \multirow{4}{8em}{\centering PL dependent \\ $\kappa_{\rm noise}$ = 1/2} 
        & \multirow{2}{4em}{\centering Parallel} & \multirow{2}{4em}{\centering $w^0L^{-1}$} & \multirow{2}{4em}{\centering $w^2L^{-1}$} & \multirow{2}{4em}{\centering $w^1L^{-1/2}$} & \multirow{2}{4em}{\centering $w^2L^{-1/2}$}  \\  &  &  & &  &   \\\cline{2-6}
    
        \multirow{2}{8em}{\centering } 
        & \multirow{2}{8em}{\centering Perpendicular}  & \multirow{2}{4em}{\centering $w^{-1}$} & \multirow{2}{4em}{\centering $w^1$} & \multirow{2}{4em}{\centering $w^{1/2}$} & \multirow{2}{4em}{\centering $w^{3/2}$}  \\  & &  &  &  &    \\ \hline

        \multirow{4}{8em}{\centering PL independent \\ $\kappa_{\rm noise}$ = 0 } 
        & \multirow{2}{4em}{\centering Parallel}  & \multirow{2}{4em}{\centering $w^{-2}L^{-2}$} & \multirow{2}{4em}{\centering $w^2L^{-2}$} & \multirow{2}{4em}{\centering $w^0L^{-1}$} & \multirow{2}{4em}{\centering $w^2L^{-1}$} \\  &  &  &  & &   \\ \cline{2-6}

        \multirow{2}{8em}{} 
        & \multirow{2}{8em}{\centering Perpendicular}   & \multirow{2}{4em}{\centering $w^{-4}$} & \multirow{2}{4em}{\centering $w^0$} & \multirow{2}{4em}{\centering $w^{-1}$}& \multirow{2}{4em}{\centering $w^1$}  \\  &  &  &  &  &  \\ \hline
     
    \end{tabular}

    \caption{Scaling laws for the sensitivity $\eta_{\rm CPW}$ as a function of the width $w$ of the CPW and the length $L$ of the probe volume along the CPW. The PL dependent noise case (first two rows) can be seen as a function of $w$ in Fig. \ref{fig:CPW}(c-f) with the saturated PL and linear PL found at small and large $w$ respectively.}
    \label{table:CPW}
\end{table*}

Comparing the two beam geometries, we see that the parallel configuration leads to better sensitivities by about two orders of magnitude compared to the perpendicular configuration, if we consider the near-saturation regime of $w\sim10\,\mu$m. This is because the parallel beam is able to interrogate a larger volume since $L$ can be arbitrarily large ($L=1$\,mm was used in the simulations), whereas the perpendicular beam is constrained by the beam ratio $L/w$ that can be practically achieved ($L/w=1$ was assumed). In practice, for dimensions $w\lesssim10\,\mu$m the parallel configuration could be achieved by patterning an optical waveguide into the diamond~\cite{Lenzini2018}. Moreover, rather than placing and aligning the patterned diamond on a pre-fabricated CPW, it may be more practical to fabricate the CPW directly on the diamond. For slope detection, a larger laser power relaxes the size requirement to obtain optimum sensitivity, from $w\sim1\,\mu$m with 10 mW of laser power to $w\sim10\,\mu$m with 1\,W. In principle, for variance detection the sensitivity keeps improving when reducing $w$, but diamond and CPW fabrication constraints are expected to impose a practical limit of $w\sim1\,\mu$m. Similarly, in the PL-independent noise regime (not shown), we can see from the scaling laws in Table \ref{table:CPW} that there will generally be an optimum width $w$ at the crossover between the linear and saturated PL regimes, i.e. at $w\sim10\,\mu$m. 

Finally, we note from Eq. \ref{ens_CPW} that the sensitivity scales with the CPW impedance as $\eta_{\rm CPW} \propto Z^{\kappa_{\rm{prot}}/2}$. This suggests that an alternate method to improve the sensitivity is by reducing the gap within the CPW. However, this becomes increasingly challenging for small $w$. For instance maintaining $Z=50\,\Omega$ with $w=10\,\mu$m requires a sub $\mu$m gap at GHz RF frequencies. Upon decreasing frequency into the MHz range, the conventional lossless transmission line approximation also becomes less applicable and the gap must be further reduced to account for the relative increase in impedance due to transitioning from an LC to RC circuit  \cite{mousavi_kinetic_2016}.

Here, we have shown the sensitivity trends with size for a single RF-spin interface geometry, defined through $\alpha_{\rm conc}$, in combination with two laser interrogation methods. These results show that a parallel laser interrogation provides greater sensitivities than perpendicular, where the detector benefits from smaller CPW track widths which enhance the sensitivity and/or reduce the laser power requirements.

\subsection{Loop antenna}

\subsubsection{Model}

\begin{figure*}
    \centering
    \includegraphics[width=1\linewidth]{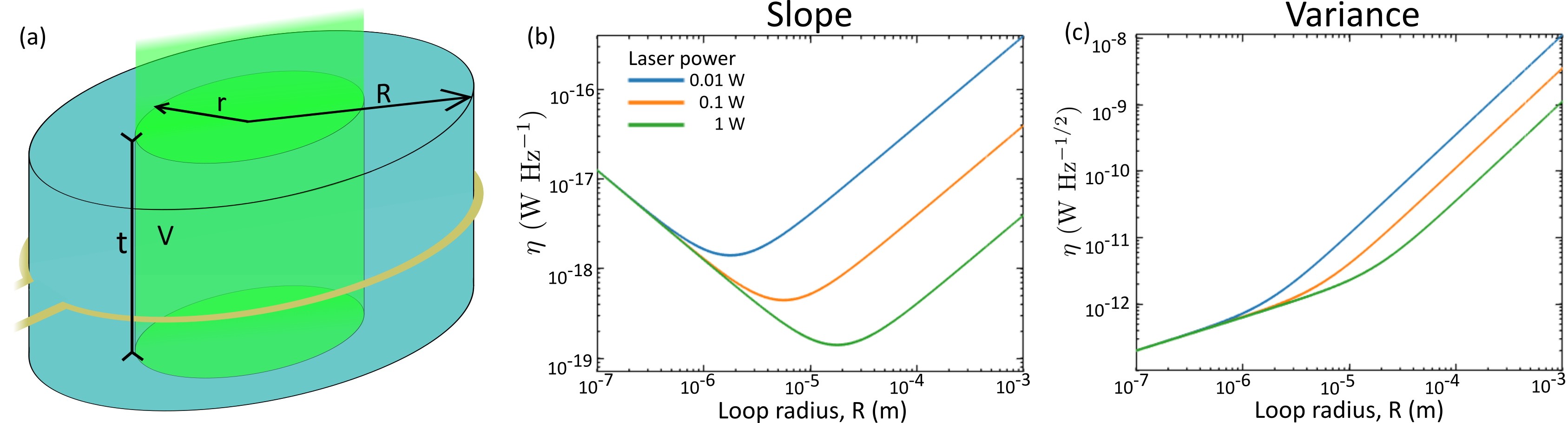}
    \caption{(a) Diagram of the loop antenna. (b)-(c) Power sensitivity of the slope (b) and variance (c) protocols as a function of the loop radius. The different curves correspond to different laser powers. The sensitivity is calculated by substituting Eq. \ref{sat} into Eq. \ref{ens_LA}. Only the PL-dependent noise regime is considered ($\kappa_{\rm noise}=1/2$). The ratio $r/R$ is taken to be 1. All other parameters are taken as per Fig. \ref{fig:CPW}.
     }
    \label{Loop_Sat_Transition}
\end{figure*}

Alternate to the CPW, the loop antenna is another common broadband solution for concentrating RF fields from kHz to GHz frequencies~\cite{Glenn2018,hermann_extending_2024}. By loop antenna here, we refer to a coaxial cable or planar waveguide terminated in a single-turn omega loop, see Fig. \ref{Loop_Sat_Transition}(a). Such systems can be modelled as a shorted transmission line where the current flowing through the loop is $I_{\rm RF}=2\sqrt{P_{\rm RF}/Z}$ with $Z=50\,\Omega$.

We assume that the diamond is contained inside the loop of radius $R$ as shown in Fig. \ref{Loop_Sat_Transition}(a). The magnetic field pointing out of plane at the centre of the coil is given by
\begin{equation}
    \label{B_LA}
    B_{\rm RF} \approx \frac{\mu_0I_{\rm RF}}{2R} = \frac{\mu_0}{R\sqrt{Z}}\sqrt{P_{\rm RF}} ~.
\end{equation}
The spatial dependence along the axial and radial direction depends on the specifications of the loop (e.g. wire diameter). In principle for a given loop geometry one would need to numerically determine the optimum probe volume that maximises the figures of merit Eqs. \ref{FOM1} and \ref{FOM2}. However, as an approximation (see Appendix \ref{app:loop num}) we can consider the optimum volume to be of radius $r=c_1R$ and thickness $t=c_2R$ commensurate to the loop radius, with fixed ratios $c_1$ and $c_2$. Moreover, we can assume the average field-to-power ratio within this optimum volume is proportional to the field at the centre of the loop
\begin{equation}
    \label{alpha_opt_LA}
    \langle\alpha_{\rm conc}^{\kappa_{\rm prot}}\rangle_V\approx \zeta\left[ \frac{\mu_0}{R\sqrt{Z}}\right]^{\kappa_{\rm prot}}~,
\end{equation}
where $\zeta$ is a constant independent of $R$.
As before, we assume that the NV spins have the correct orientation to maximise coupling to the out-of-plane magnetic field for the sensing protocol considered.

Substituting Eq. \ref{alpha_opt_LA} and the volume $V=\pi r^2t=\pi c_1^2c_2R^3$  into Eq. \ref{ens}, we obtain the power sensitivity equation for the loop antenna,
\begin{equation}
    \label{ens_LA}
    \eta_{\rm loop} = \left[\frac{\xi_{\rm noise}}{\zeta\left[ \frac{\mu_0}{R\sqrt{Z}}\right]^{\kappa_{\rm prot}}\beta_{\rm prot}^{\kappa_{\rm prot}}(\rho_{\rm PL}\pi c_1^2c_2R^3)^{1-\kappa_{\rm noise}}}\right]^{2/{\kappa_{\rm prot}}}~.
\end{equation}

\begin{table*}
    \begin{tabular}{ |c|c|c|c|c| } 
         \hline
        \multirow{4}{8em}{\centering Noise regime} &   \multicolumn{4}{c|}{$\eta_{loop}$}   \\ \cline{2-5}   & \multicolumn{2}{c|}{\centering Slope,  $\kappa_{\rm prot} = 1$}  & \multicolumn{2}{c|}{\centering Variance,  $\kappa_{\rm prot} = 2$} \\ \cline{2-5} &    \multirow{2}{8em}{\centering Saturated PL} &  \multirow{2}{8em}{\centering Linear PL} &  \multirow{2}{8em}{\centering Saturated PL}  & \multirow{2}{8em}{\centering Linear PL}   \\ &   &  & &  \\
        \hline
    
        \multirow{3}{8em}{\centering PL dependent \\ $\kappa_{\rm noise}$ = 1/2} 
         & \multirow{3}{4em}{\centering $R^{-1}$} & \multirow{3}{4em}{\centering $R^1$} & \multirow{3}{4em}{\centering $R^{1/2}$} & \multirow{3}{4em}{\centering $R^{3/2}$}  \\  &   & &  &  \\  &   & &  &  \\\cline{1-5}

        \multirow{3}{8em}{\centering PL independent \\ $\kappa_{\rm noise}$ = 0 } 
        &   \multirow{3}{4em}{\centering $R^{-4}$} & \multirow{3}{4em}{\centering $R^0$} & \multirow{3}{4em}{\centering $R^{-1}$} & \multirow{3}{4em}{\centering $R^{1}$} \\  &  &  & & \\  &  &  &  &   \\ \cline{1-5}

    \end{tabular}

    \caption{Scaling laws for the sensitivity $\eta_{\rm loop}$ as a function of the loop radius $R$. The PL-dependent noise case (first row) can be seen as a function of $R$ in Fig. \ref{Loop_Sat_Transition}(b)-(c) with the saturated PL and linear PL found at small and large $R$ respectively.
    }
    \label{table:LA}
\end{table*}

\subsubsection{Scaling laws}

Similar to the CPW case, from Eq. \ref{ens_LA} we can derive scaling laws here as a function of the loop radius $R$. In the saturated PL regime, we obtain
\begin{equation}
    \label{scaling_LA}
    \eta_{\rm loop} \propto R^{2-6\frac{1-\kappa_{\rm noise}}{\kappa_{\rm prot}}}~. 
\end{equation}
In the linear regime and considering a laser beam perpendicular to the plane of the loop as shown in Fig. \ref{Loop_Sat_Transition}(a), we obtain 
\begin{equation}
    \label{scaling_LA2}
    \eta_{\rm loop} \propto R^{2-2\frac{1-\kappa_{\rm noise}}{\kappa_{\rm prot}}}~. 
\end{equation}
The explicit scaling laws in the different scenarios considered are given in Table \ref{table:LA}.

\subsubsection{PL-dependent noise regime}

We now discuss the $R$ dependence of the sensitivity in the PL-dependent noise regime ($\kappa_{\rm noise}=1/2$). Fig. \ref{Loop_Sat_Transition}(b)-(c) plots $\eta_{\rm loop}$ calculated from Eq. \ref{ens_LA} when substituting the general saturation law for $\rho_{\rm PL}$ in Eq. \ref{sat} for both the slope and variance detection cases. Similarly to the CPW, large $R$ corresponds to the linear PL regime, whereas small $R$ corresponds to the saturated PL regime.

The trends are identical to the CPW with the perpendicular beam. Namely, for slope detection there is an optimum radius at the cross-over between the linear and saturated regimes ($R\sim10\,\mu$m for 0.1W), while for variance detection the sensitivity keeps improving as $R$ is reduced with a decrease in slope after the transition. In practice, such small loops may be realised by fabricating a planar structure on a substrate or even directly on the diamond, either as a terminated waveguide or as an in-line $\Omega$-shaped structure~\cite{Opaluch2021,Li2022}. The thickness $t$ of the probe volume would then be controlled by the thickness of the NV layer. 
\subsection{Geometry comparison}

To compare the sensitivity of the CPW and loop antenna, we compute the ratio of Eqs. \ref{ens_CPW} and \ref{ens_LA} in the saturated PL regime,
\begin{equation}
    \label{CPW_loop}
    \frac{\eta_{\rm CPW}}{\eta_{\rm loop}} \sim \frac{w^{2-4\frac{1-\kappa_{\rm noise}}{\kappa_{\rm prot}}}L^{-2\frac{1-\kappa_{\rm noise}}{\kappa_{\rm prot}}}}{R^{2-6\frac{1-\kappa_{\rm noise}}{\kappa_{\rm prot}}}}~,
\end{equation}
where we ignored factors of the order of unity for simplicity. 
For the purpose of comparison, we consider the cross-over point where $w\sim R\sim10\,\mu$m representing an optimum in some cases, which gives
\begin{equation}
    \label{CPW_loop_Parallel}
    \frac{\eta_{\rm CPW}}{\eta_{\rm loop}}   \sim \left(\frac{R}{L}\right)^{2\frac{1-\kappa_{\rm noise}}{\kappa_{\rm prot}}}.
\end{equation}
In the parallel laser beam geometry for the CPW, the length $L$ is completely independent and so can be taken to be of the order of 1\,mm (limited by available diamond size), yielding a ratio $L/R \sim100$. In the PL-dependent noise regime, this implies that the CPW is more sensitive than the loop antenna by a factor $\sim100$ for slope detection and $\sim10$ for variance detection, and even more with PL-independent noise. This highlights the advantage of a waveguide geometry which confines the magnetic field in two dimensions only, leaving the third dimension available for maximising the probe volume.

\section{Absolute sensitivity estimates} \label{sec:num}

We now discuss the absolute values of power sensitivities predicted by our analysis, using realistic parameters. We consider the photon shot noise limit, which sets $\kappa_{\rm noise}=1/2$ and $\xi_{\rm{noise}} =1 $. 

The NV-related parameters are chosen as follows. The NV density is assumed to be $\rho_{\rm{NV}} = 8\times10^{23}~{\rm m}^{-3}$ equivalent to 4.5 ppm, corresponding to typical commercially available NV-doped diamonds. For most sensing protocols (e.g. those based on dynamical decoupling sequences), the field-to-contrast ratio is roughly given (for slope detection) by $\beta_{\rm{prot}}\sim C_{\rm max}\gamma_e \tau$ where $C_{\rm max}$ is the maximum optical contrast, $\gamma_e=28$~GHz/T is the electron's gyromagnetic ratio, and $\tau$ is the evolution time of the sequence~\cite{Degen2017}. We take $\beta_{\rm{prot}}=10^{4}~{\rm T}^{-1}$ approximately corresponding to $C_{\rm max}\sim 3\%$ and $\tau\sim 10~\mu$s, compatible with the spin coherence time (under dynamical decoupling) of a diamond with 4.5 ppm of NVs. The saturation excitation intensity of the NV is taken to be $I_{\rm sat}=1$\,mW$/\mu$m$^2$~\cite{Manson2006}. The collected PL rate per NV at saturation, $\sigma_{\rm NV}$, is approximated here by the product of the PL emission rate at saturation ($\sim 10^{-7}~{\rm s}^{-1}$), the laser duty cycle of the sequence ($10\%$, assuming 1-$\mu$s laser pulses and an evolution time of $\tau\sim 10~\mu$s), and the collection efficiency (10\%), resulting in $\sigma_{\rm NV}=10^{5}~{\rm s}^{-1}$.

Since we are looking for rough estimates, we take the factors of order unity to be exactly 1 for simplicity, i.e. $\zeta=c_1=c_2=1$. Moreover, for the CPW we take $Z = 50~\Omega$ and $L = 1$~mm for the parallel excitation geometry (or $L=w$ in the case of perpendicular excitation). The above parameters were used to generate the curves in Fig. \ref{fig:CPW} and \ref{Loop_Sat_Transition}. We note that apart from the saturation intensity $I_{\rm sat}$ which dictates the location of the crossover point between linear and saturated regimes, the other parameters only change the values on the vertical axes, i.e. the trends as a function of track width or loop radius are unchanged otherwise. 

As discussed previously, the best sensitivity among the geometries analysed is achieved for the CPW with parallel excitation and a track width of order $w\sim10~\mu$m (with a laser power of 1 W). The corresponding power sensitivity is found to be about $10^{-20}$ W Hz$^{-1}$ for slope detection, and $10^{-12}$ W Hz$^{-1/2}$ for variance detection. The value of $10^{-20}$ W Hz$^{-1}$ for slope detection is comparable to the sensitivity of state-of-the-art conventional (software defined radio) RF detectors, which operate equivalently to a slope detection mode. For reference, the corresponding magnetic sensitivities found by multiplying by the factor $\alpha_{\rm conc}$ are 1 pT Hz$^{-1/2}$ and (10 nT)$^2$ Hz$^{-1/2}$, respectively. These magnetic sensitivities are in line with previously reported experimental values, highlighting that our power sensitivity estimates are realistic.

\section{Conclusion}

We theoretically analysed the power sensitivity of NV-based RF detectors and found it follows optimisation procedures that differ from conventional magnetometry. We found power sensitivity often has a preference for smaller detector sizes over larger, maximising both the concentrated magnetic field and the interrogation rate from fixed intensity laser sources. This theory enables insight to system design for detecting weak RF signals through the advantageous use of miniaturisation to increase sensitivity and lower laser power requirements. 

The best geometry among those analysed in this paper was the coplanar waveguide with a parallel laser beam excitation. With an optimum CPW track width of $w\sim10\,\mu$m, we predict shot-noise-limited power sensitivities of $10^{-20}$ W Hz$^{-1}$ (slope detection) and $10^{-12}$ W Hz$^{-1/2}$ (variance detection) may be achievable.   

Even better sensitivities may potentially be achieved using resonant structures rather than the broadband RF concentrators analysed here. In particular, several types of resonators have been studied for driving the NV Larmor frequency typically around 3\,GHz, such as the planar ring antenna~\cite{Sasaki2016}, loop gap resonator~\cite{Eisenach2018}, and high-permittivity dielectric resonator~ \cite{vallabhapurapu_fast_2021}. Resonance frequencies in the low MHz range would lead to significantly larger structures, making effective use of resonators more challenging for lower frequency NV-sensing devices. In a resonator, the magnetic field for a given input power is enhanced at the resonance, which may lead to larger maximum values for the field-to-power ratio $\alpha_{\rm conc}$. However, the geometry of a resonator (and hence the probed volume) is dictated by the resonance condition, and so further analysis will be required to see if RF resonators can provide a net sensitivity benefit. We stress that this would come at the cost of a narrow operation bandwidth (most likely in the GHz range) while the broadband structures analysed here can detect RF signals from kHz to GHz. Moreover, most RF sensing protocols require a microwave driving field at the NV Larmor frequency in addition to the RF test signal, which makes the use of resonators challenging.

Even for broadband structures, there are physical limitations that should also be considered when engineering detection systems that we do not rigorously address in this work, such as reducing RF concentrator size often leads to increased difficulty with maintaining low impedance. This reduces the ratio $\alpha_{\rm{conc}}$ in the small size regime, and therefore reduces the optimum sensitivity, which may be investigated through further theoretical analysis. Regardless of these limitations, the trends presented in this paper represent a useful guide and starting point for the experimental realisations of optimised NV-based RF detectors.

\section*{Acknowledgements}
This work was supported by the Australian Government under the Advanced Strategic Capabilities Accelerator through its Emerging and Disruptive Technologies (EDT) Program (in partnership with Diamond Defence Pty Ltd. and Phasor Innovation Pty Ltd.). N. G. acknowledges support by the Defence Science Institute, an initiative of the State Government of Victoria. D. A. B. acknowledges support from the Australian Research Council through grant DE230100192.






\newpage
\bibliography{bibliography}

\appendix

\begin{figure*}
    \centering
    \includegraphics[width=1\linewidth]{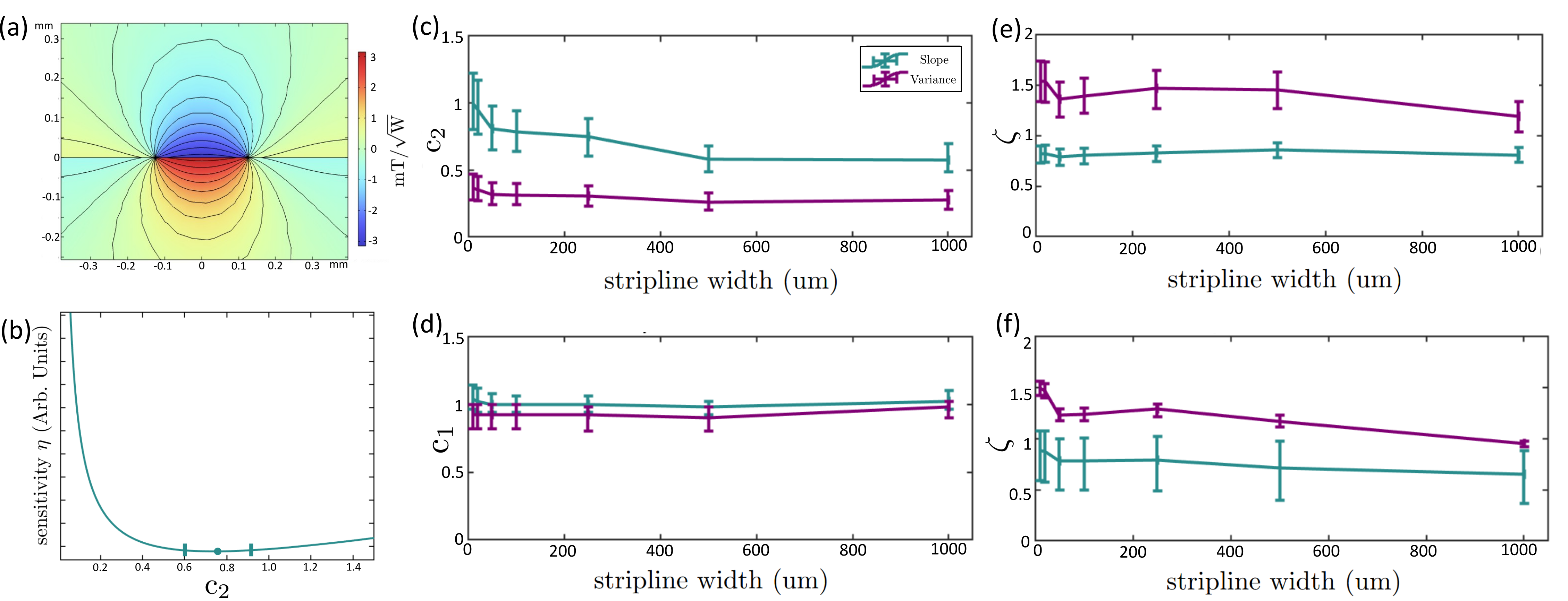}
    \caption{(a) COMSOL simulation example of a 250 $\mu$m stripline, with a gap of 5 $\mu$m at 100 MHz. The colour map depicts the x-component of the magnetic field. (b) Sensitivity scaling at fixed $c_1$ = 0.95 for a stripline width of 250 $\mu$m, showing the optimum $c_2$ value (circle) and the definition of error bars used in (c)-(f) plots marked either side, and set at 1\%. c (d) Optimal $c_2$ [$c_1$] values for fixed $c_1$ = 0.95 [$c_{2}$: slope = 0.8, variance = 0.4] with varying stripline widths. (e) [(f)] zeta values from fixed $c_1$ [$c_2$] values present in (c) [(d)].}
    \label{fig:CPW_Numeric}
\end{figure*}

\section{CPW: Numerical simulations}
\label{app:CPW num}

In the analysis, we made simplifying assumptions about the optimum $c_1$, $c_2$ and the average magnetic field enclosed in their defined region. Here we perform numerical simulations using COMSOL Multiphysics to validate these assumptions. The model as shown in Fig. \ref{fig:CPW_Numeric}(a) involves a two dimensional modal analysis with a 200 nm thick gold stripline. The colour map depicts the horizontal component of the RF magnetic field contributing to the contrast.

First, we numerically determine the values of $c_1$ and $c_2$ that maximise the product $\langle\alpha_{\rm conc}^{\kappa_{\rm prot}}\rangle_V V^{1-\kappa_{\rm noise}}$, and hence minimise the sensitivity according to Eq. \ref{ens} (see an example optimisation in Fig. \ref{fig:CPW_Numeric}(b)), as a function of the CPW width, $w$. These simulations are carried out with constant impedance within 1\% tolerance.  The result of the optimisation is shown in Fig. \ref{fig:CPW_Numeric}(c)-(d). We see that the optimum $c_2$  for slope and variance are 0.8 and 0.4 respectively, whilst the optimum $c_1$ is around 0.95 and relatively independent of $w$. Numerical optimisation of the $c$ values in the linear regime, where the figure of merit is $\langle\alpha_{\rm conc}^{\kappa_{\rm prot}}\rangle_V d^{1-\kappa_{\rm noise}}$, is not presented as they trivially require reduction until close to saturation to yield better sensitivities.

Next, we used these optimised values of $c_1$ and $c_2$ and compute $\langle\alpha_{\rm conc}^{\kappa_{\rm prot}}\rangle_V$ over this optimised probed region. This is plotted as a function of $w$ in Fig. \ref{fig:CPW_Numeric}(e)-(f) in terms of the ratio $\zeta$ to the expected maximum magnetic field value given by Eq. \ref{B_CPW}. We see that for both $\kappa_{\rm prot}=1,2$ the factor $\zeta$ is relatively independent of $w$ with a value around 0.8 and 1.5.

A mild trend of decreasing $\zeta$ can be seen as stripline width increases, which we believe to be best described by edge current concentrating due to skin and proximity effects. At smaller wavelength to stripline width ratios, magnetic field localisation around the gaps of the coplanar waveguide can occur. This increases the field density at the edges whilst simultaneously decreasing the uniformity across the stripline. This effectively causes a ratio change between $c_2$ and $\zeta$ values that decrease the optimal $c_2$. This is inadequately compensated with an increase in $\zeta$ that generally fails to maintain similar sensitivities given by similar striplines without these concentrated edge currents. This furthermore encourages decreasing sensor size in the high frequency RF regime to minimise these effects.

\section{Loop antenna: Numerical simulations}
\label{app:loop num}

\begin{figure*}
    \centering
    \includegraphics[width=1\linewidth]{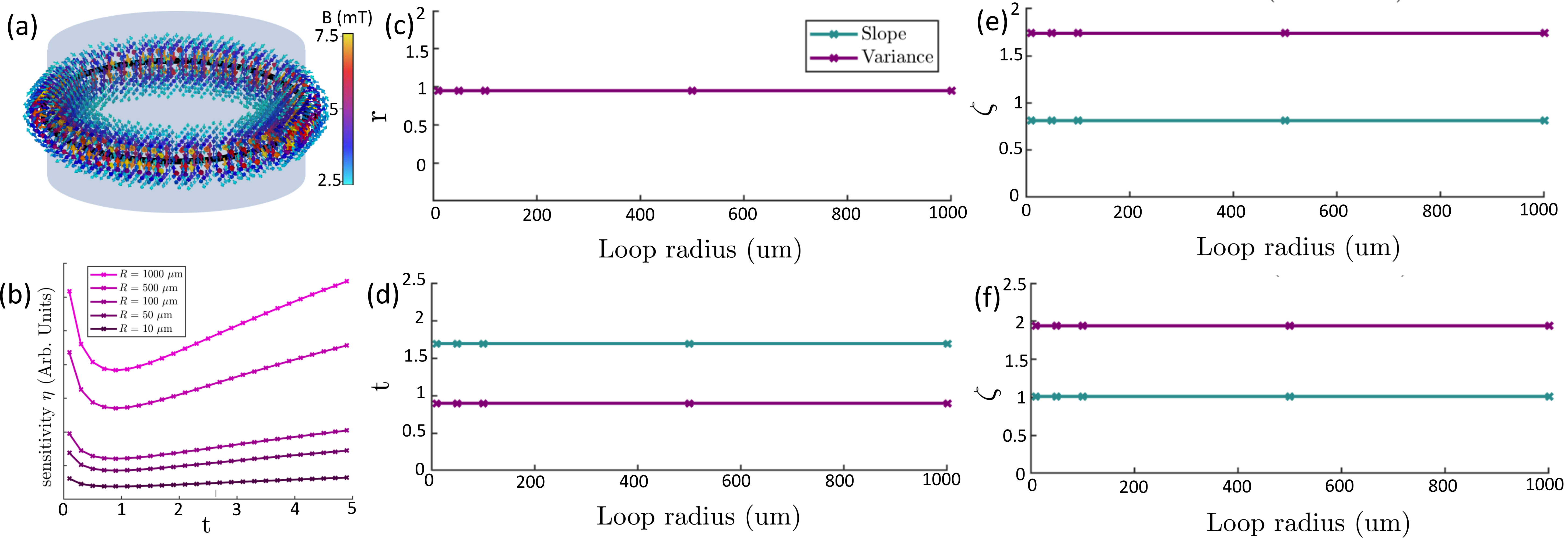}
    \caption{(a) Visualisation of loop antenna CST simulations with magnetic field strengths less than 2.5 mT omitted. (b) sensitivity scaling of variance at a  fixed r = 0.85R for varying loop antenna radii, R. (c) [(d)] optimal r [t] values for fixed t$_{slope/var}$ = 1.7R/0.9R  [r = 0.85R] with varying loop antenna radii, R. (e) [(f)] zeta values from fixed t [r] values present in (c) [(d)].}
    \label{fig:LA_Numeric}
\end{figure*}

Similarly to the coplanar waveguide we numerically simulate the loop antenna to verify the optimal parameters and the conversion factor scaling. We perform these using CST Studio Suite, modelling the loop in three dimensions. We vary loop diameter whilst keeping the conductor diameter proportional to loop radius, with a 3 MHz RF. The input and output lumped ports are present at a small gap in the loop as seen in Fig. \ref{fig:LA_Numeric}a. The cylindrical coordinate based mesh lacks constant density so we approximate volume with a 3D circular segment. First we find the optimal thickness, $t$, and probe radius, $r$, for the laser at different loop radiuses that maximise $\zeta$ as seen in Fig \ref{fig:LA_Numeric}(c-d), where we omit error bars present in the CPW simulations due to lack of resolution. 

Using these values we calculate their corresponding $\zeta$ values. The CST loop antenna simulations were not designed to keep constant impedance, so to enable direct comparison between the varying loop radii we normalise the relative magnetic strength by the central field magnitude, effectively normalising out the impedance through Eq. \ref{B_LA}. These results as seen in Fig. \ref{fig:LA_Numeric}(e-f) show an independence of $\zeta$ from $R$ as expected, supporting the assumptions behind our derivation. These results also show a $\zeta$ value comparable to or greater than 1, reflective of our assumption that the maximum field is located in the centre of the loop, when it is shown to exist near the wires.

\section{Loop antenna: Case of fixed diamond thickness}

It can be undesirable to make custom  diamonds for each particular experiment and instead one may seek to make the best RF detector out of the diamond currently available, e.g. for a fixed NV layer thickness. Here we analyse the loop antenna again but assuming a fixed diamond thickness, T, removing the degree of freedom present in the previous optimisation. The saturated regime then scales as 

\begin{equation}
    \label{scaling_LA_T}
    \eta_{\rm loop}  \propto \zeta~^{2/\kappa_{\rm prot}} R^{2-4\frac{1-\kappa_{\rm noise}}{\kappa_{\rm prot}}}T^{-2\frac{1-\kappa_{\rm noise}}{\kappa_{\rm prot}}}~
\end{equation}
and the linear regime as 
\begin{equation}
    \label{scaling_LA_T_Linear}
    \eta_{\rm loop} \propto \zeta~^{\kappa_{\rm prot}/2} R^{2}T^{-2\frac{1-\kappa_{\rm noise}}{\kappa_{\rm prot}}}~. 
\end{equation}
The scaling of $\zeta$ is however no longer trivially constant, as the probe region is no longer just confined to the optimal volume. To approximate this affect we take the magnetic field about the centre line passing through the loop from the Biot-Savart law
\begin{equation}
    B_Z = \frac{\mu_0}{4 \pi} \frac{2 \pi R^2I}{(z^2 + R^2)^{3/2}}
\end{equation}
and use these values as a rough model for the change in field strength for the whole vertical plane. For the slope protocol, after integrating over the thickness of the diamond and using the definition of $\zeta = B_{avg}/B_{max}$ we find the average field across the diamond to scale approximately by
\begin{equation}
    \zeta \propto \frac{R}{\sqrt{R^2+T^2}}~. 
\end{equation}
We can generalise to include the variance-detection case leading to the general form of
\begin{equation}
    \zeta_{\kappa_{\rm{prot}}} \propto (\frac{R}{\sqrt{R^2+T^2}})^{\kappa_{\rm{prot}}}~. 
\end{equation}

This relation is then applied to Eq. \ref{scaling_LA_T} and \ref{scaling_LA_T_Linear} giving the saturated regime scaling as 

\begin{equation}
    \label{scaling_LA_T}
    \eta_{\rm loop}  \propto (R^2+T^2) R^{-4\frac{1-\kappa_{\rm noise}}{\kappa_{\rm prot}}}T^{-2\frac{1-\kappa_{\rm noise}}{\kappa_{\rm prot}}}~
\end{equation}
and the linear regime as 
\begin{equation}
    \label{scaling_LA_T_Linear}
    \eta_{\rm loop} \propto  (R^2+T^2) T^{-2\frac{1-\kappa_{\rm noise}}{\kappa_{\rm prot}}}~. 
\end{equation}

Considering only the noise dependent regime, these relations show that for a fixed diamond thickness and a non-saturating laser power, smaller loop size is still preferential. Furthermore in the saturated regime, variance based sensitivity reaches a horizontal plateau as $R \rightarrow 0$ whilst slope has a minima (optimal) sensitivity at $R\sim T$.

\section{Summary of symbols used in the manuscript} \label{sec:SumTable}

Table \ref{Appx:SumTable} lists the symbols used in this manuscript along with their definitions and units. Additionally, where applicable, the numerical values used to generate the sensitivity plots in Fig. \ref{fig:CPW} and \ref{Loop_Sat_Transition} are indicated.

\begin{table*}

    \begin{tabular}{ |C|J|L|L| }
         \hline
        Symbol & \Tstrut Definition\Bstrut & Units & Value assumed \\
        \hline
        \hline
        $P_{\rm RF}$ & \Tstrut Input RF power\Bstrut & W & - \\
        \hline
        $B_{\rm RF}(\vec{r})$ & \Tstrut RMS amplitude of the magnetic field (projected on the relevant axis) at position $\vec{r}$\Bstrut & T & - \\
        \hline
        $\alpha_{\rm conc}(\vec{r})$ & \Tstrut Factor relating RF power to magnetic field according to $B_{\rm RF}(\vec{r})=\alpha_{\rm conc}(\vec{r})\sqrt{P_{\rm RF}}$\Bstrut & T\,W$^{-1/2}$ & - \\
        \hline
        $C(\vec{r})$ & \Tstrut Relative NV PL change (contrast) induced by the RF\Bstrut & - & - \\
        \hline
        $\kappa_{\rm prot}$ & \Tstrut Exponent indicating variance or slope detection, such that $C\propto B_{\rm RF}^{\kappa_{\rm prot}}$\Bstrut & - & 2 for variance, 1 for slope \\
        \hline 
        $\beta_{\rm prot}$ & Factor relating contrast to magnetic field according to $C(\vec{r})=[\beta_{\rm prot}B_{\rm RF}(\vec{r})]^{\kappa_{\rm prot}}$  & T$^{-1}$ &\Tstrut $10^{4}~{\rm T}^{-1}$ \Bstrut  \\  
        \hline
        $\eta_{\rm single}(\vec{r})$ & Power sensitivity for a single NV & \Tstrut$~$ W Hz$^{-1/2}$ for variance,$~~$ W Hz$^{-1}$ for slope\Bstrut & - \\
        \hline
        $I_{\rm PL}$ & \Tstrut Collected photon rate\Bstrut & s$^{-1}$ & - \\ 
        \hline
        $t$ & \Tstrut Total measurement time\Bstrut & s & - \\
        \hline
        $\kappa_{\rm noise}$ & \Tstrut Exponent characterising the noise scaling, such that the fluctuation in the photon number is $N\propto I_{\rm PL}^{\kappa_{\rm noise}}\sqrt{t}$\Bstrut & - & 0 for PL-independent noise, 0.5 for PL-dependent noise \\
        \hline
        $\xi_{\rm noise}$ & \Tstrut Pre-factor accounting for the magnitude of the noise such that $N=\xi_{\rm noise}I_{\rm PL}^{\kappa_{\rm noise}}\sqrt{t}$\Bstrut & - & 1 in the photon shot noise limit \\
        \hline
        $\eta_{\rm ens}$ & \Tstrut Ensemble-averaged power sensitivity for a given volume of NVs\Bstrut & $~~$ W Hz$^{-1/2}$ for variance, W Hz$^{-1}$ for slope & - \\
        \hline
        $V$ & \Tstrut Interrogation volume\Bstrut & m$^{3}$ & - \\
        \hline
        $\rho_{\rm PL}$ &\Tstrut PL rate per unit volume (within the interrogation volume)\Bstrut & s$^{-1}$ m$^{-3}$ & -  \\
        \hline
        $\langle x\rangle_V$ & \Tstrut Spatial average of $x$ over the probed volume\Bstrut & same as $x$ & - \\
        \hline
        $\rho_{\rm NV}$ & \Tstrut NV density\Bstrut & m$^{-3}$ & $8\times10^{23}~{\rm m}^{-3}$ \\
        \hline
        $P_{\rm laser}$ &\Tstrut Laser power\Bstrut & W & - \\
        \hline
        $A$ & \Tstrut Cross-sectional area of the laser beam\Bstrut & m$^2$ & - \\
        \hline
        $I_{\rm sat}$ & \Tstrut Laser intensity at NV saturation\Bstrut & W m$^{-2}$ & $10^9$ W m$^{-2}$  \\
        \hline
        $\sigma_{\rm NV}$ & \Tstrut Collected PL rate per NV at saturation\Bstrut & s$^{-1}$ & $10^{5}~{\rm s}^{-1}$ \\
        \hline
        ${\rm FoM}_{\rm sat}$ & \Tstrut Geometry-dependent figure of merit in the saturation regime\Bstrut & (T\,W$^{-1/2})^{\kappa_{\rm prot}}{\rm m}^{3(1-\kappa_{\rm noise})}$ &  \\
        \hline
        ${\rm FoM}_{\rm lin}$ & \Tstrut Geometry-dependent figure of merit in the linear regime\Bstrut & (T\,W$^{-1/2})^{\kappa_{\rm prot}}{\rm m}^{1-\kappa_{\rm noise}}$ & - \\
        \hline
        $d$ & \Tstrut Distance traversed by the laser beam across the probe volume\Bstrut & m & - \\
        \hline
        
    \end{tabular}
\end{table*}

\begin{table*}

    \begin{tabular}{ |C|J|L|L| }

     \hline
            Symbol & \Tstrut Definition\Bstrut & Units & Value assumed \\
        \hline
        \hline
        $\gamma_e$ & \Tstrut Electron's gyromagnetic ratio\Bstrut & GHz/T & 28\,GHz/T \\
        \hline
        $C_{\rm max}$ & \Tstrut Maximum optical contrast\Bstrut & - & $\sim3\%$ \\
        \hline
        $\tau$ & \Tstrut Evolution time of the sensing sequence\Bstrut & s & $\sim10~\mu$s \\
        \hline
        $w$ & \Tstrut CPW track width\Bstrut & m & - \\
        \hline
        $Z$ & \Tstrut Transmission line impedance\Bstrut  & $\Omega$ & $50\,\Omega$ \\
        \hline
        $I_{\rm RF}$ & \Tstrut RMS current in the transmission line\Bstrut & A & -  \\
        \hline
        $\mu_0$ & \Tstrut Vacuum permeability\Bstrut & H/m & 4$\pi\times 10^{-7}$ H/m \\
        \hline
        $c_1$ & \Tstrut Normalised width of the probe volume (absolute width is $c_1w$ or $c_1R$)\Bstrut & - & 1  \\
        \hline
        $c_2$ & \Tstrut Normalised height of the probe volume (absolute height is $c_2w$ or $c_2R$)\Bstrut & - & 1  \\
        \hline
        $L$ & \Tstrut Length of the probe volume\Bstrut & m & 1 mm  \\
        \hline
        $R$ & \Tstrut Radius of the loop\Bstrut & m & -  \\
        \hline
        $r$ &\Tstrut Radius of the probe volume\Bstrut & m & -  \\
        \hline
        $t$ & \Tstrut Thickness of the probe volume\Bstrut & m & -  \\
        \hline
        $\zeta$ & \Tstrut Ratio between approximated magnetic field and actual average magnetic field over the interrogation volume\Bstrut & - & 1\\
        \hline
    \end{tabular}
        \caption{Summary table of the symbols used to derive the sensitivity formulas present in this paper. The last column indicates the values assumed to generate the sensitivity plots in Fig. \ref{fig:CPW} and \ref{Loop_Sat_Transition}.}
        \label{Appx:SumTable}
\end{table*}

\end{document}